\documentclass{qjmam}
\usepackage{amssymb}
\usepackage{amsmath}
\graphicspath{{./figures/}}
\usepackage{cleveref}
\usepackage{color}
\usepackage{calrsfs}
\usepackage[symbol]{footmisc}

\startpage{1}
\yr{2025}

\newcommand{\pfr}[2]{\ensuremath{\frac{\partial #1}{\partial #2}}}

\newcommand\Rey{\mbox{\textit{Re}}}

\newcommand{\beq}{\begin{equation}}
\newcommand{\eeq}{\end{equation}}

\DeclareMathAlphabet\mathbfcal{OMS}{cmsy}{b}{n}

\DeclareMathAlphabet\mathbit
    \encodingdefault\rmdefault\bfdefault\itdefault
\DeclareOldFontCommand{\bi}{\normalfont\bfseries\itshape}{\mathbit}

\newcommand{\be}{\begin{equation}}
\newcommand{\ee}{\end{equation}}

\def\fakebold#1{\relax\ifvmode\leavevmode\fi%
\ifmmode%
\setbox0=\hbox{$#1$}%
\else%
\setbox0=\hbox{#1}%
\fi%
\kern-.02em\copy0 \kern-\wd0%
\kern .04em\copy0 \kern-\wd0%
\kern-.0125em\raise.02em\box0%
}%

\begin{document}

\title[{Kerr--Dold vortices in an axisymmetric stagnation point flow}] {Kerr--Dold vortices in an axisymmetric stagnation point flow}

\author[P.~Rajamanickam] {Prabakaran Rajamanickam}

\address{Department of Mathematics and Statistics, University of Strathclyde, Glasgow G1 1XQ, UK}

\received{\recd 7 June 2025}

\maketitle

\eqnobysec

\begin{abstract} 
The existence of Kerr--Dold-type vortices in axisymmetric stagnation point flow is demonstrated, extending the class of known thick vortex solutions.
\end{abstract}

\section{Introduction}
\label{sec:intro}

Stretched vortex lines or surfaces are ubiquitous in real-life flow fields, particularly in turbulent flows~\cite{moffatt1994stretched}. Among these, stretched vortex tubes are especially common, as evidenced by tube-like vorticity contours observed in many direct numerical simulations. In contrast, vortex surfaces or sheets tend to be unstable due to the Kelvin–Helmholtz instability. The vortex structures are described by a Burgers-type mechanism involving vortex stretching, diffusion, and convection. Two canonical configurations of such vortices are well known: the classical Burgers vortex in an axially straining flow, and the two-dimensional Burgers vortex sheet in a planar stagnation point flow. Burgers vortices are generally stable solutions of the Navier–-Stokes equations~\cite{leibovich1981global,robinson1984stability,prochazka1995two,crowdy1998note,schmid2004three}, whereas the Burgers vortex sheet becomes unstable above a critical Reynolds number~\cite{aryshev1981stability,lagnado1984stability,beronov1996linear}. Additionally, Burgers vortex solutions have been identified in radial stagnation point flows as well~\cite{rajamanickam2021steady}.

Standard Burgers-type vortices, including those in non-axisymmetric straining fields~\cite{prochazka1998structure}, typically exhibit exponentially decaying vortex tails and are therefore classified as  \textit{thin vortex structures}. However, stagnation point flows can also support vortex structures with algebraically decaying tails, which can be termed \textit{thick vortex structures}.  In 1994, Kerr and Dold~\cite{kerr1994periodic} discovered a family of solutions consisting of a periodic array of counter-rotating thick stretched vortices in planar stagnation point flows~\cite{drazin2006navier}. Experimental observations of these Kerr–Dold vortices have been reported in Taylor’s four-roll mill apparatus~\cite{lagnado1990visualization,andreotti2001experiment}, in crossed rectangular channels~\cite{kalashnikov1991ordered} and in microfluidic crossed channels~\cite{burshtein2019controlled,burshtein2021periodic}. Notably, Kerr~\cite{kerr2024periodicsteadyvorticesstagnation} has shown that a wide variety of Kerr–Dold vortices can be constructed by manipulating the amplitude and phase of the vortex tails; see also~\cite{towns1999roll}. The Kerr--Dold vortices within a Hiemenz flow has been studied in~\cite{xiong2004distortion}. Recent studies~\cite{migdal2021asymmetric,shariff2021viscous} have also introduced thick vortex sheet solutions, further expanding the landscape of thick vortex structures; the solutions reported in~\cite{bazant2005exact} are likely to be thick vortex structures as well.

Interestingly, the existence of thick Kerr–Dold-type vortex solutions has not yet been explored in axisymmetric stagnation point flows. The objective of this short paper is to demonstrate that such solutions can indeed be constructed in this setting. Rather than providing an exhaustive catalogue of solutions, this paper focuses on showcasing a few numerically computed examples.

\section{Governing equations and boundary conditions}

Consider the axisymmetric stagnation point flow expressed in cylindrical coordinates $(r^*,\theta,z^*)$. The velocity components are given by
\begin{equation}
    v_r^*= -kr^* , \qquad v_\theta^*=0, \qquad v_z^* = 2kz^* \label{radial}
\end{equation}
where $k>0$ denotes the strain rate. Hereafter, we non-dimensionalise the problem by scaling time with $1/k$ and lengths with $\sqrt{\nu/k}$, where $\nu$ is the kinematic viscosity of the fluid.

Let us look for vortex solutions superposed on the axisymmetric stagnation flow~\eqref{radial}. Specifically, we write the velocity and vorticity components as
\begin{equation}
    \begin{bmatrix}
-r\\
0\\
2z
\end{bmatrix}+
 \Rey\begin{bmatrix}
u(r,\theta)\\
v(r,\theta)\\
0
\end{bmatrix} 
\qquad \text{and} \qquad 
     \Rey\begin{bmatrix}
0\\
0\\
\omega(r,\theta)
\end{bmatrix} \label{decomp}
\end{equation}
where $\Rey$ can be interpreted as a vortex Reynolds number. We focus on periodic solutions with  fundamental period $2\pi/N$ where $N=2,3,4,\dots$, excluding $N=1$ for reasons that will become clear shortly. These solutions may describe a pair of counter-rotating vortices within the angular interval $\theta\in[-\pi/N,+\pi/N]$.

The continuity equation can be satisfied automatically when we introduce the vortex streamfunction $\psi(r,\theta)$ such that
\begin{equation}
    u= \frac{1}{r}\pfr{\psi}{\theta}, \quad v=-\pfr{\psi}{r}.
\end{equation}
The required governing equations then reduce to
\begin{align}
    -\omega &= \frac{1}{r}\pfr{}{r}\left(r\pfr{\psi}{r}\right) + \frac{1}{r^2}\pfr{^2\psi}{\theta^2}, \label{psi}\\
   \frac{\Rey}{r}\pfr{(\omega,\psi)}{(r,\theta)} - r\pfr{\omega}{r} - 2 \omega &=\frac{1}{r}\pfr{}{r}\left(r\pfr{\omega}{r}\right) + \frac{1}{r^2}\pfr{^2\omega}{\theta^2}. \label{omega}
\end{align}
In the azimuthal direction, periodic boundary conditions are enforced,
\begin{equation}
    \psi(r,\theta-\pi/N)=\psi(r,\theta+\pi/N), \qquad \omega(r,\theta-\pi/N)=\omega(r,\theta+\pi/N), \label{periodic}
\end{equation}
whereas, in the radial direction, it is sufficient to impose homogeneous Dirichlet boundary conditions,
\begin{equation}
   u,v,\omega\to 0 \quad \text{as}\,\,r\to 0 \quad \text{and} \quad \text{as}\,\,r\to \infty. \label{BC}
\end{equation}
In fact, as $r\to 0$ and as $r\to\infty$, the nonlinear terms in~\eqref{omega} will become negligible and the asymptotic behaviour is dictated by the linearised equations. These are discussed in the next section, from which we can deduce that
\begin{align}
     \omega\sim r^N,\quad \psi \sim r^N,\quad u \sim r^{N-1}, \quad v\sim r^{N-1} \quad \text{as}\quad r\to 0,
\end{align}
and
\begin{equation}
   \omega \sim r^{-2}, \quad \psi-f(\theta)\sim r^{-2}, \quad
       u\sim r^{-1}, \quad v \sim r^{-3}  \quad \text{as}\quad r\to \infty \label{BCinfinity}
\end{equation}
where $f(\theta)$ is a periodic function. 

The exclusion of the $N=1$ case is now evident since it imply non-zero velocities at the symmetry axis $r=0$, which is unphysical.  Furthermore, the $r^{-2}$ decay of vorticity at large $r$ causes the (non-dimensional) circulation $\Gamma=\Gamma^*/\nu$ encircling an individual vortex in the pair to diverge logarithmically, i.e.,
\begin{equation}
    \lim_{R\to \infty}\int_{0}^{\pi/N} \int_0^R \omega \,rdrd\theta  \sim \ln R. \label{Regamma}
\end{equation}
Thus, the Reynolds number $\Rey$ cannot be directly linked to the circulation $\Gamma$. Instead, we can introduce, for example, a normalization condition such as
\begin{equation}
    \frac{N}{2} \int_0^{\pi/N}\int_0^\infty\frac{\omega}{r} \,rdrd\theta = 1, \label{norm}
\end{equation}
which serves to fix the definition of the vortex Reynolds number.

\begin{figure}[h!]
\centering
\includegraphics[width=0.45\textwidth]{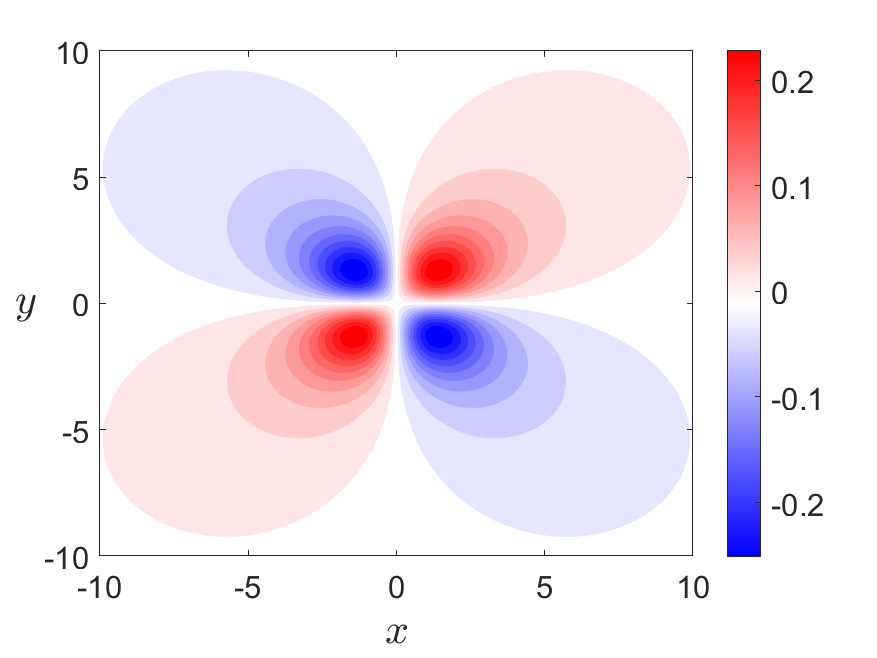}
\includegraphics[width=0.45\textwidth]{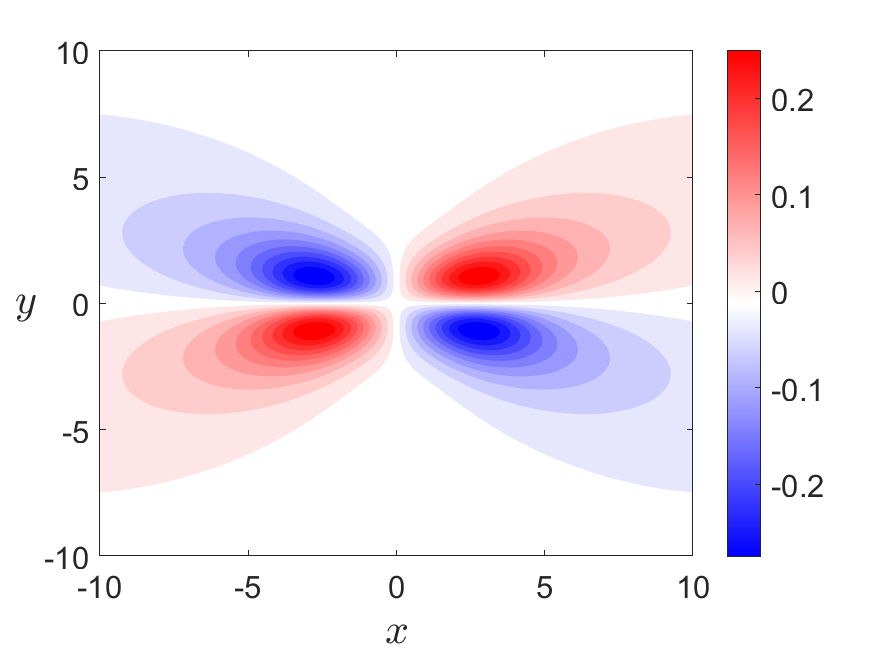}
\includegraphics[width=0.45\textwidth]{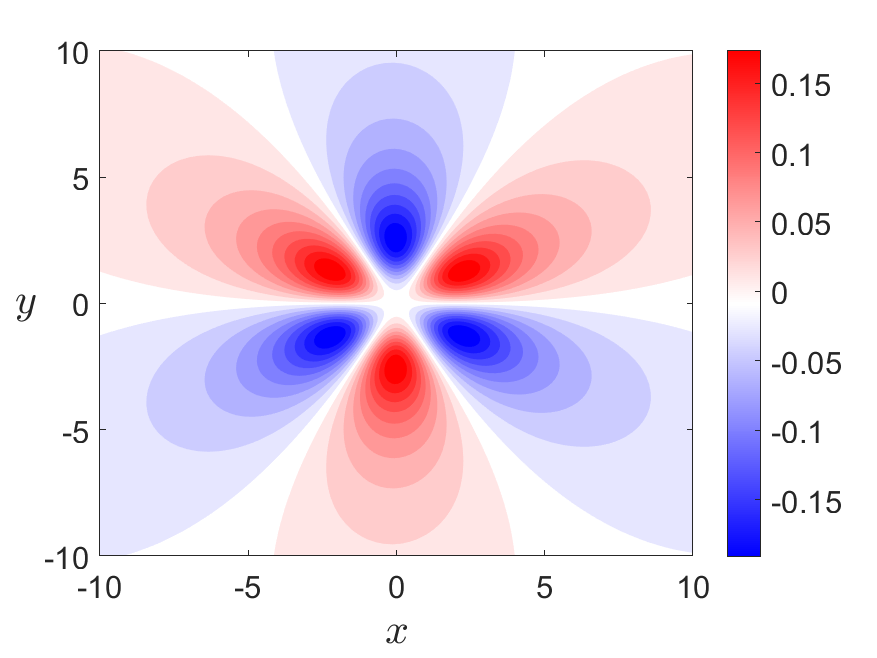}
\includegraphics[width=0.45\textwidth]{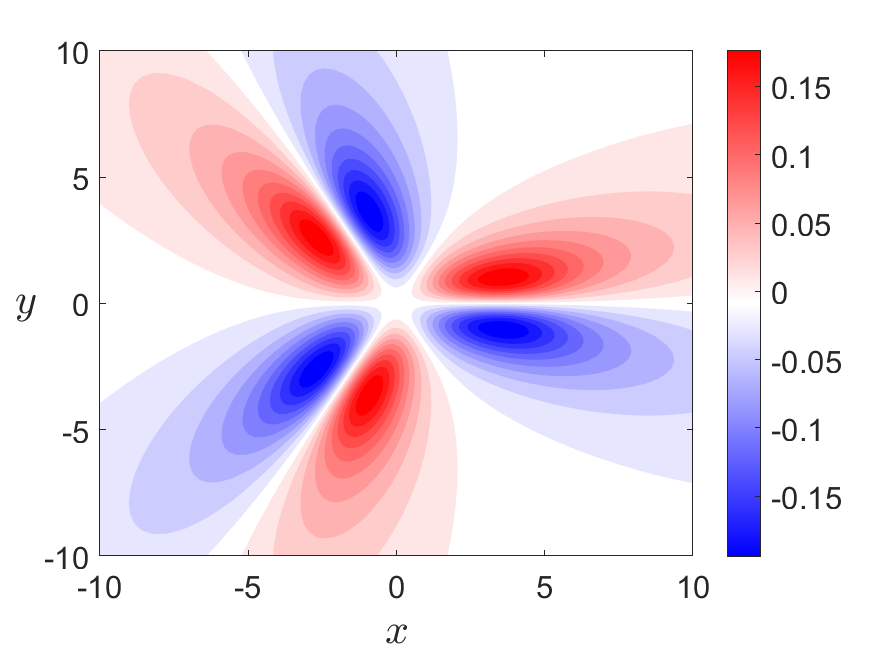}
\includegraphics[width=0.45\textwidth]{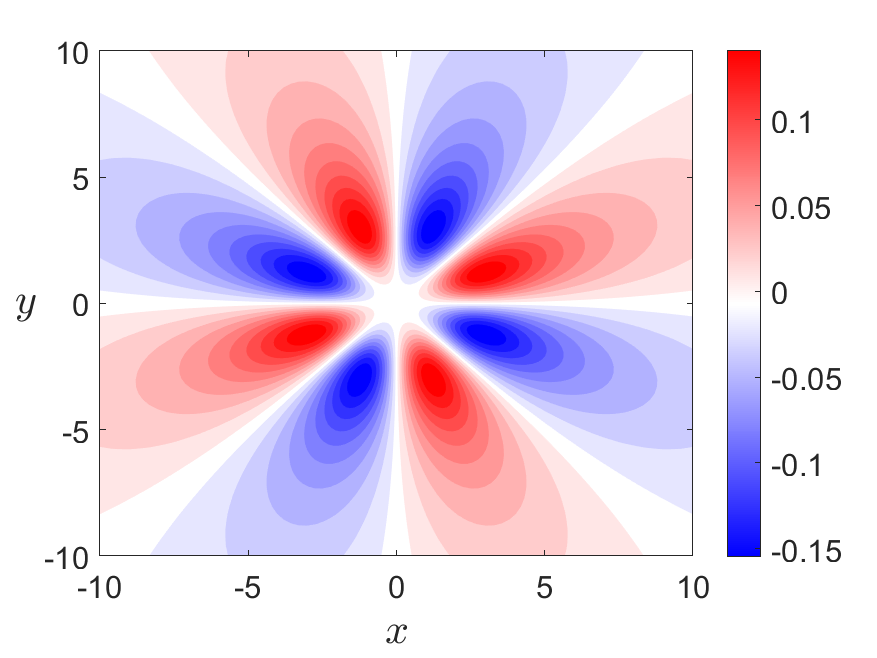}
\includegraphics[width=0.45\textwidth]{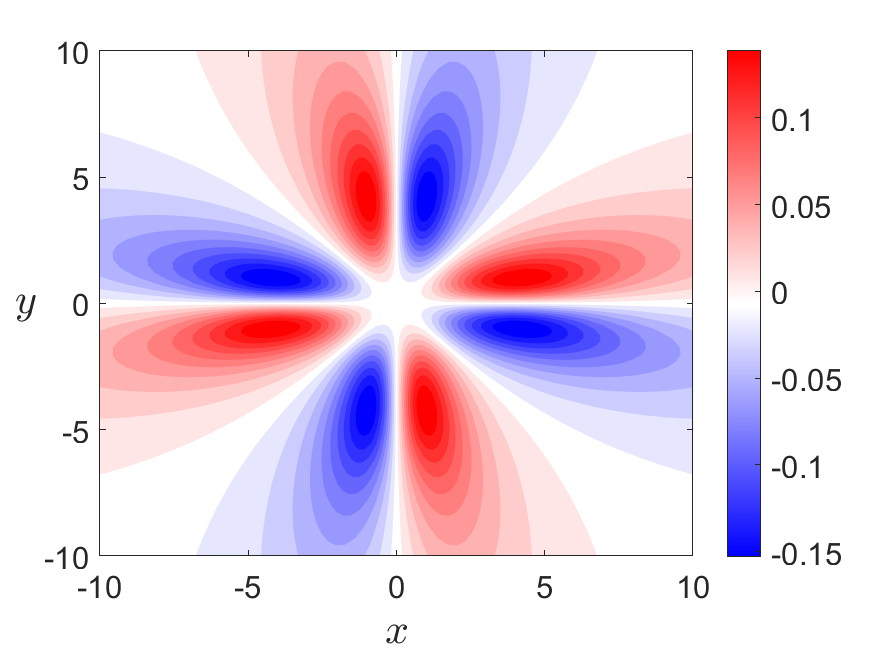}
\caption{Illustrative numerical results showing vorticity contours: The top row corresponds to $\Rey=1$ and $10$ for $N=2$, the middle row to $\Rey=1$ and $20$ for $N=3$ and the bottom row to $\Rey=1$ and $30$ for $N=4$.} 
\label{fig:vortex}
\end{figure}

\section{Solution for small vortex Reynolds numbers}

For small values of the vortex Reynolds numbers $\Rey$, the nonlinear term in~\eqref{omega} is negligible, and the problem reduces to a linear system at leading order. Given the periodicity of the solution in the azimuthal direction, it is natural to represent the vorticity and streamfunction using Fourier series expansions,
\begin{equation}
    \omega(r,\theta)=\frac{1}{2}\sum_{n=1}^\infty\left[\hat\omega_n e^{inN\theta}+ \overline{\hat\omega_n}e^{-inN\theta}\right], \qquad  \psi(r,\theta)=\frac{1}{2}\sum_{n=1}^\infty\left[\hat\psi_n e^{inN\theta}+ \overline{\hat\psi_n}e^{-inN\theta}\right]
\end{equation}
where $\hat\omega_n=\hat\omega_n(r)$ and $\hat\psi_n=\hat\psi_n(r)$ are complex-valued functions of $r$. The axisymmetric mode $n=0$ is omitted  since it corresponds either to the  Burgers vortex solution ($\hat\omega_0\sim e^{-r^2/2}$), or to its companion solution ($\hat\omega_0\sim e^{-r^2/2}\mathrm{Ei}(r^2/2)$) which  diverges logarithmically at $r=0$. 

Since the problem is linear, the Fourier modes decouple, each satisfying the system of equations
\begin{equation}
    \hat\omega_n''+ \left(r+\frac{1}{r}\right)\hat\omega_n' + \left(2-\frac{n^2N^2}{r^2}\right)\hat\omega_n = 0, \qquad \hat\psi_n''+\frac{1}{r}\hat\psi_n'-\frac{n^2N^2}{r^2}\hat\psi_n=-\hat\omega_n.
\end{equation}
Among the two linearly independent solutions of the homogeneous equation for $\hat\omega_n$, the one that remains regular at the origin is given by
\begin{equation}
    \hat\omega_n(r)=\frac{A_n}{2} re^{-r^2/4}\left[I_{\frac{nN-1}{2}}(r^2/4) - I_{\frac{nN+1}{2}}(r^2/4)\right]  \label{omegaasymp}
\end{equation}
where $I$ denotes the modified Bessel function of the first kind and $A_n=\int_0^\infty \hat\omega_n dr$ is an arbitrary complex constant that fixes the vortex amplitude of the $n$-th Fourier mode. Note that we assume $A_1\neq 0$ since the fundamental periodicity requires the first harmonic mode $n=1$ to be non-trivial.

The  homogeneous solutions to the equation for $\hat \psi_n$ are $r^{nN}$ and $r^{-nN}$.  Using the method of variation of parameters and enforcing the boundary conditions that velocity components vanish as $r$ approaches zero and infinity, we obtain
\begin{align}
    \hat\psi_n(r) = \frac{r^{nN}}{2nN}\int_r^\infty s^{1-nN}\hat\omega_n(s)\,ds  + 
     \frac{r^{-nN}}{2nN} \int_0^r s^{1+nN}\hat\omega_n(s)\,ds. \label{psiquad}
\end{align}
Applying standard integral identities for modified Bessel functions, the above expression can be simplified to the closed form,
\begin{equation}
    \hat\psi_n(r) = \frac{A_n}{2nN} re^{-r^2/4}\left[I_{\frac{nN-1}{2}}(r^2/4) + I_{\frac{nN+1}{2}}(r^2/4)\right].
\end{equation}
Since $\hat \psi_n\to A_n\sqrt{2/\pi}/nN+ O(1/r^2)$ as $r\to\infty$, we have
\begin{equation}
    f(\theta)= \frac{1}{\sqrt{2\pi}}\sum_{n=1}^\infty\frac{1}{nN}\left[A_ne^{inN\theta}+\overline{A_n}e^{-inN\theta}\right],
\end{equation}
thereby defining the streamfunction value as $r$ approaches infinity (see~\eqref{BC}) in terms of vortex amplitude coefficients $A_n$. 

The above construction yields a family of infinitely many solutions, parametrised by the sequence of vortex amplitudes $\{A_n\}$ with $A_1\neq 0$, or equivalently by the form of the periodic function $f(\theta)$. These linear solutions can be continued numerically to finite Reynolds numbers. Representative numerical results for the branch corresponding to $A_1=-i$ and $A_2=A_3=\cdots=0$ are illustrated in Figure~\ref{fig:vortex}.

\section*{Acknowledgements}

The author would like to thank Adam D. Weiss for the many enjoyable and engaging conversations about the problem discussed in this work.

\bibliographystyle{unsrt}
\bibliography{references}

\end{document}